# A Critical Analysis of Affirmative Action and Academic Performance: A Response to the Preprint Authored by Matheus et al. (2025)

Claudio Andre Barbosa de Lira[1*] and Ricardo Borges Viana[2]

[1]Faculty of Physical Education and Dance, Federal University of Goiás (Goiânia, Brazil)

[2]Institute of Physical Education and Sports, Federal University of Ceará (Fortaleza, Brazil)

**\*Corresponding author:**

Claudio Andre Barbosa de Lira, PhD

claudioandre@ufg.br

**Abstract**

This article offers a critical response to the preprint by Matheus et al. (2025), which evaluates the academic performance of students admitted through different entry routes at São Paulo State University. Although the dataset compiled by the authors is valuable, their analysis contains conceptual, methodological, and historical limitations that undermine the validity of their conclusions. We argue that the preprint misinterprets the purpose of Brazil's affirmative action policy, which is intended to ensure equitable access rather than equalize academic performance at the point of university entry. Early performance differences among universal system, public school, and racial-quota students reflect longstanding inequalities in Brazil's basic education system. Furthermore, the preprint generalizes findings basically from only three undergraduate programs (Physics, Biology, and Pedagogy) to represent entire academic domains (STEM, Biological Sciences, and Humanities), an extrapolation that reduces the reliability of its conclusions. Overall, we contend that early performance differences indicate structural educational inequality, not policy failure. Affirmative action fulfills its mandate by guaranteeing access, while universities bear the responsibility for providing equitable conditions for academic progression.

## 1. Introduction

We read with great interest the preprint by Matheus et al. [1] evaluating the academic performance of students admitted through different entry routes at São Paulo State University (UNESP). The authors conclude that students admitted through racial quotas (Black, Brown, and Indigenous quota *[Pretos, Pardos e Indígenas]* - PPI) perform worst, students from public schools (EP) occupy an intermediate position, and students admitted through universal system (SU) perform best. While the dataset compiled by the authors is potentially valuable, their analytical framework and interpretive approach contain conceptual, methodological, and historical limitations that prevent their conclusions from being sustained.

What follows is not a defense of affirmative action as an ideological position but a defense of analytical rigor. A policy cannot be evaluated when its objectives are misunderstood, its mechanisms are mischaracterized, or its structural antecedents are ignored. We aim to clarify these issues so the debate remains empirically grounded and socially responsible.

It is important to emphasize that the central focus of Brazil's quota system is public-school origin rather than racial or ethnic identity alone. Although the authors acknowledge that the PPI quota is a subcategory within the public-school quota, a comprehensive analysis must compare students admitted through the SU with those admitted through the public-school route, which includes both PPI and non-PPI students. Public schools in Brazil have historically operated under suboptimal learning conditions, including limited financial resources, overcrowded classrooms, inadequate laboratory facilities, and disparities in teacher training [2–5]. These structural limitations affect students' academic preparation and make public-school background the primary dimension that affirmative action policies seek to address. Assessing quota effectiveness without situating the analysis within this broader context risks overlooking the systemic inequalities the policy is meant to mitigate.

## 2. Structural Inequality Precedes Admission and Cannot Be Ignored

Differences in early performance among SU, EP, and PPI students are predictable consequences of Brazil's deeply unequal basic education system. These differences stem from disparities in school quality, language instruction, laboratory access, extracurricular exposure, cultural capital, family income, and available study time. Such inequalities are well documented in the literature and precede university admission [5,6].

Affirmative action is not designed to equalize academic performance at entry; it is designed to equalize access. Academic performance is shaped by what occurs after students are admitted, including the quality of teaching, availability of institutional support, sense of academic belonging, material conditions, and access to pedagogical resources. For clarity, we reproduce UNESP's official description of its affirmative action objective (translated from institutional documents) [7]: "The University Council of UNESP developed an inclusion proposal to fulfill its social responsibility and ensure access to its undergraduate programs for students from public schools and for Black, Brown, and Indigenous students." This institutional statement clarifies that the policy addresses access, not immediate academic parity.

Expecting affirmative action to eliminate post-admission performance disparities misinterprets both its purpose and its operational boundaries. The leveling of academic outcomes depends on a distinct set of institutional commitments, including tutoring programs, bridge courses, mentoring structures, financial assistance, and inclusive pedagogical strategies. These mechanisms occur after admission and are separate from admissions policies. Therefore, disparities observed during the early stages of university education should be attributed to the long-standing inequalities students bring with them and to the effectiveness of university-level support, rather than to an affirmative action policy whose mandate is limited to ensuring equitable entry.

## 3. Conceptual Misunderstanding: Preuniversity Inequality Is Not an Effect of Affirmative Action

Central to the authors' interpretation is the assumption that performance differences among universal system (*Sistema Universal* [SU]), public-school (*Escola Pública* [EP]), and racial-quota (*Preto, Pardo e Indígena* [PPI]) students can be used to evaluate the effectiveness of UNESP's affirmative action policy. Conceptually, this assumption is incorrect.

UNESP's admissions process internally reorders candidates so that high-performing quota applicants are shifted into the SU category. As a result, the SU group is not composed solely of "non-quota" students. Rather, it consists of the highest-scoring applicants across all admission routes. The preprint therefore compares groups with systematically different pre-university conditions and attributes these differences to a policy that, by design, does not eliminate preexisting inequalities at the point of admission.

The interpretive flaw is conceptual rather than statistical. The authors attribute to the quota system what is, in reality, a predictable consequence of unequal access to high-quality basic education, language instruction, extracurricular opportunities, laboratory infrastructure, cultural capital, and available study time. In doing so, the preprint confuses the effects of structural inequality with the effects of the policy intended to mitigate that inequality.

## 4. Lack of Representativeness: Three Programs Cannot Represent Three Knowledge Domains

The authors analyzed Biology and Pedagogy undergraduate programs and then generalized their findings to the broader fields of Biological Sciences and the Humanities, respectively. This extrapolation is methodologically indefensible.

UNESP is a large multicampus university offering dozens of distinct undergraduate programs across these domains [8]. Table 1 illustrates the breadth of available programs. Asserting that Pedagogy "represents the Humanities" or that Biology "represents Biological Sciences" lacks empirical justification and introduces an avoidable generalization error.

Regarding science, technology, engineering, and mathematics (STEM), the authors examined success rates in Calculus I across all UNESP programs that include the course in their curricula. Although this analysis is less biased and more representative, it still overlooks the fact that Calculus I may not serve as a comprehensive or reliable proxy for performance across STEM fields. Further details are discussed in the next section.

**Table 1.** Undergraduate Courses offered by São Paulo State University (UNESP)

| Biological Sciences | STEM | Humanities |
|---|---|---|
| Biological Sciences | Computer Science | Administration |
| Biomedical Sciences | Aeronautical Engineering | Public Administration |
| Ecology | Environmental Engineering | Architecture and Urbanism |
| Physical Education | Cartographic Engineering and Surveying | Archival Science |
| Nursing | Civil Engineering | Theatre Arts/Performing Arts |
| Agronomic Engineering | Food Engineering | Visual Arts |
| Fisheries Engineering | Bioprocess and Biotechnology Engineering | Library Science |
| Forest Engineering | Biosystems Engineering | Economics |
| Pharmacy | Control and Automation Engineering | Social Sciences |
| Physiotherapy | Energy Engineering | Journalism |
| Speech-language Pathology and Audiology | Materials Engineering | Communication: Radio, TV, and Internet |
| Medicine | Production Engineering | Design |
| Veterinary Medicine | Electrical Engineering | Law |
| Nutrition | Electronic and Telecommunications Engineering | Philosophy |
| Dentistry | Wood Industrial Engineering | Geography |
| Occupational Therapy | Mechanical Engineering | History |
| Animal Science | Chemical Engineering | Languages and Literature |
| | Statistics | Languages and Literature –Translation |
| | Physics | Music |
| | Medical Physics | Pedagogy |

|  | Geology | Pedagogy (Distance Learning) |
|---|---|---|
|  | Mathematics | Psychology |
|  | Meteorology | International Relations |
|  | Chemistry | Public Relations |
|  | Information Systems | Social Work |
|  |  | Tourism |
|  |  | Business Administration |

Note: STEM refers to science, technology, engineering, and mathematics.

5. **The Importance of Longitudinal Patterns: Convergence Is Possible and Empirically Observed**

A fundamental limitation of the preprint is its exclusive focus on initial courses such as Calculus I, Cell Biology, and Philosophy of Education I. Differences in early performance between students from unequal educational backgrounds are expected. However, longitudinal analyses are necessary to determine whether these gaps persist over time, and such analyses frequently reveal patterns of convergence.

Preliminary observations from a Physical Education program at a Federal University of Ceará (data not yet published) demonstrate substantial grade differences between universal system and affirmative-action students in a second-semester Gross Human Anatomy course. Yet these differences disappear in later courses such as Human Physiology, Exercise Physiology, Kinesiology, and Biomechanics. This pattern suggests that, when adequate instructional conditions are provided, academic convergence is not only possible but likely.

Regarding STEM, the authors assessed success rates in a mid-program course (Electromagnetism I), but they restricted this evaluation to Physics students alone. As a result, the analysis cannot be considered representative of STEM as a whole.

Thus, evaluating a decade-long policy primarily based on early-semester performance fundamentally misrepresents the Biological Sciences, STEM, and Humanities programs at UNESP. Using early-course grades as the main criterion for judging the policy is as misleading as evaluating a medical intervention based only on the patient's state at admission.

6. **Unsupported Insinuations About Academic Rigor in Pedagogy**

The authors state: "[…] It would be important also for researchers in the Humanities to propose appropriate metrics for the evaluation of student performance, as approval in early courses or the overall graduation rates are making all students appear equally good." Interpreted straightforwardly, this implies that passing or failing courses in Pedagogy may not be the primary mechanism by which academic quality is determined. Such a claim effectively questions the legitimacy of an entire discipline's evaluation practices. Any assertion of this magnitude requires robust evidence, which the preprint does not provide. If unsupported, it should be withdrawn to avoid reinforcing stereotypes.

Moreover, by suggesting that conventional academic indicators in Pedagogy fail to distinguish student performance, the authors inadvertently cast doubt on the rigor and credibility of the field without offering a substantive analytical basis. This stance disregards the complex, multidimensional competencies that Pedagogy programs are designed to cultivate, competencies that extend beyond simplistic assumptions about grading leniency. Without presenting empirical evidence demonstrating systematic flaws in evaluation standards, the authors risk perpetuating dismissive narratives toward fields associated with teaching and social care, narratives that have contributed to their marginalization within the Brazilian academic hierarchy.

## 7. Evidence Contradicts the Narrative of a "Reputation Threat"

At the end of the preprint, the authors express concern about safeguarding UNESP's institutional reputation. Yet their own data indicate high retention rates even among SU and EP students. For example, in Biology, the program the authors treat as their most representative case, approximately 52% of SU and EP students graduate within 8 years. Considering that the minimum program duration is 4 or 5 years, depending on shift (full-time or single shift), this completion rate demonstrates that those who graduate do so by meeting all academic standards of one of Brazil's most prestigious universities.

Graduation from UNESP requires fulfilling rigorous academic criteria, and these standards apply equally to all students regardless of entry route.

Furthermore, to address the authors' claim of reputational risk, we examined two robust Brazilian indicators: the Folha de São Paulo University Ranking (*Ranking Universitário da Folha de São Paulo* [RUF]) (see section 7.1) and the General Course Index (*Índice Geral de Cursos* [IGC]) from Brazil's Ministry of Education (see section 7.2).

**7.1 RUF position (2012–2025)**

UNESP has consistently ranked among the top 10 universities in Brazil since the creation of the Folha de São Paulo University Ranking. This long-term stability provides no evidence of reputational decline following the implementation of affirmative action policies. Instead, it suggests institutional continuity and resilience across the entire period evaluated (Table 2).

**Table 2**. Position of UNESP in the Folha de São Paulo University Ranking from 2012 to 2025.

| Year | Ranking | Link: |
|---|---|---|
| 2025 | 6 of 204 | https://ruf.folha.uol.com.br/2025/ranking-de-universidades/principal/ |
| 2024 | 6 of 203 | https://ruf.folha.uol.com.br/2024/ranking-de-universidades/principal/ |
| 2023 | 6 of 197 | https://ruf.folha.uol.com.br/2023/ |
| 2022 | Not disclosed | Not applicable |
| 2021 | Not disclosed | Not applicable |
| 2020 | Not disclosed | Not applicable |
| 2019 | 6 of 197 | https://ruf.folha.uol.com.br/2019/ranking-de-universidades/principal/ |
| 2018 | 8 of 196 | https://ruf.folha.uol.com.br/2018/ranking-de-universidades/ |
| 2017 | 7 of 195 | https://ruf.folha.uol.com.br/2017/ranking-de-universidades/ |
| 2016 | 6 of 195 | https://ruf.folha.uol.com.br/2016/ranking-de-universidades/ |
| 2015 | 6 of 192 | https://ruf.folha.uol.com.br/2015/ranking-de-universidades/ |
| 2014 | 6 of 192 | https://ruf.folha.uol.com.br/2014/rankingdeuniversidades/ |
| 2013 | 6 of 192 | https://ruf.folha.uol.com.br/2013/ |
| 2012 | 6 of 188 | https://ruf.folha.uol.com.br/2012/ |

**7.2 IGC scores (2007–2023)**

The IGC evaluates higher education institutions based on undergraduate performance, National Student Performance Exam (*Exame Nacional de Desempenho de Estudantes* [ENADE]) scores, and postgraduate performance indicators from the Coordination for the Improvement of Higher Education Personnel (*Coordenação de Aperfeiçoamento de Pessoal de Nível Superior* [CAPES]) [9,10]. The IGC scale ranges from 1 (lowest) to 5 (highest).

UNESP reached its highest IGC scores (grade 5) after implementing affirmative action policies. While these results do not allow for causal inference, they contradict any assertion that institutional quality or academic performance declined during or after policy implementation (Table 3).

**Table 3:** General Course Index (IGC) of UNESP.

| Year | IGC |
|---|---|
| 2023 | 5 |
| 2022 | 5 |
| 2021 | Not disclosed |
| 2020 | Not disclosed |
| 2019 | 5 |
| 2018 | 5 |
| 2017 | 5 |
| 2016 | 4 |
| 2015 | 4 |
| 2014 | 5 |
| 2013 | 4 |
| 2012 | 4 |
| 2011 | 4 |
| 2010 | 4 |
| 2009 | 4 |
| 2008 | 4 |
| 2007 | 4 |

Note: Data from the electronic database of the Brazilian Ministry of Education [10].

## 8. Ethical inconsistencies

The manuscript does not indicate whether the study was reviewed or approved by an Institutional Ethics Committee. Because the authors analyzed identifiable, nonpublic academic records from a university information system, prior ethics approval is mandatory under Brazilian research guidelines. Conducting research involving sensitive student data without explicit oversight raises serious ethical concerns and may constitute non-compliance with research governance standards.

## 9. Conclusion

Affirmative action does not fail when quota students initially obtain lower grades. It succeeds by fulfilling its primary function: ensuring equitable access to higher education for historically excluded groups. What constitutes institutional failure is the absence of adequate pedagogical, material, and psychosocial support needed to guarantee equitable academic progression after admission.

The analyses presented by Matheus et al. offer useful descriptive insights but require interpretation within appropriate conceptual and historical frameworks. Their preprint attributes to the affirmative action policy outcomes that stem from deep and persistent structural inequalities, inequalities that long precede a student's university entry and that higher education institutions have a responsibility to mitigate rather than reproduce.

Brazil lived under slavery for more than three centuries. A frequently cited metaphor observes that if the entire history of Brazil were represented as one hour, approximately 45 minutes would have been lived under slavery [11]. This is not an exaggeration for rhetorical effect. It underscores that racial inequality is structurally embedded in Brazilian society.

Interpreting early academic performance disparities as evidence that affirmative action "harms the university" disregards this historical context. The relevant question is not whether quota students begin their studies from different academic starting points. They do, because Brazilian society systematically produces unequal opportunities. The question that universities must confront is whether they are willing to provide the institutional conditions necessary to foster equity, academic success, and belonging for all admitted students.